\documentclass[11pt]{article} 
\usepackage{epsfig}

\renewcommand{\vec}[1]{\ensuremath{\mathbf {#1}}}
\newcommand{\qnvec}[1]{\ensuremath{\underline {#1}}}

\newcommand{\qcomm}[2]{\ensuremath{\left[#1,#2\right]}}

\newcommand{\qp}{\ensuremath{ \,\,.}}
\newcommand{\qk}{\ensuremath{ \,\,,}}

\newcommand{\qmean}[1]{\ensuremath{\overline {#1}}}
\newcommand{\qom}[1]{{\omega_{\vec #1}}}
\newcommand{\qons}[1]{{#1^+}}

\newcommand{\qVd}{\ensuremath{{\cal H}}}
\newcommand{\qbra}[1]{\ensuremath{\left<#1\right|}}
\newcommand{\qket}[1]{\ensuremath{\left|#1\right>}}

\newcommand{\qexv}[1]{\ensuremath{\left<#1\right>}}

\begin{document}
\begin{flushright} \small
  TUW--03--19 \\ hep-th/0306101
\end{flushright}
\bigskip

\begin{center}
  {\large\bfseries Time ordered perturbation theory for non-local
    interactions;
    \\applications to NCQFT} \\[5mm]
  \renewcommand{\thefootnote}{\fnsymbol{footnote}} S.
  Denk${^1}$\footnote[1]{Work supported by DOC [predoc program of the
    \"Osterreichische Akademie der Wissenschaften].}
  and M. Schweda$^2$
  \footnote[7]{Work supported by the FWF: P15015, P15463.} \\[3mm]
  {\small\slshape Institute for Theoretical Physics, Vienna University
    of
    Technology,
    \\ Wiedner Hauptstrasse 8--10, A-1040 Vienna, Austria \\[5 mm]
    $^1${\upshape\ttfamily denk@hep.itp.tuwien.ac.at}\\[3mm]
    $^2${\upshape\ttfamily mschweda@tph.tuwien.ac.at}}
\end{center}
\vspace{5mm} \renewcommand{\thefootnote}{\arabic{footnote}}
\setcounter{footnote}{1} \hrule\bigskip

\centerline{\bfseries Abstract} \medskip In the past decades, time
ordered perturbation theory was very successful in describing
relativistic scattering processes. It was developed for local quantum
field theories. However, there are field theories which are governed
by non-local interactions, for example non-commutative quantum field
theory (NCQFT).  Filk \cite{Filk96} first studied NCQFT perturbatively
obtaining the usual Feynman propagator and additional phase factors as
the basic elements of perturbation theory. However, this treatment is
only applicable for cases, where the deformation of space-time does
not involve time.  Thus, we generalize Filk's approach in two ways:
First, we study non-local interactions of a very general type able to
embed NCQFT. And second, we also include the case, where non-locality
involves time.  A few applications of the obtained formalism will also
be discussed.

\bigskip\hrule\bigskip
%
%
%
%%%%%%%%%%%%%%%%%%%%%%%%%%%%%%%%%%%%%%%%%%%%%%%%%%%
\section{Introduction}
%%%%%%%%%%%%%%%%%%%%%%%%%%%%%%%%%%%%%%%%%%%%%%%%%%%
%
%
Quantum field theory (QFT) on non-commutative space-time shows
completely unexpected and fascinating features in the perturbative
realization of the various non-commutative field theories like the
ordinary scalar field theory invoked as a toy model, the gauge-field
models (with and without supersymmetry) etc.  Especially, the
non-planar part of the tadpole graph of a scalar NCQFT in
4-dimensional space-time shows that the non-commutative phases of the
four point interaction create an UV-finite result connected with a new
type of IR-singularity for vanishing external momenta (Naive power
counting arguments would lead to the conclusion that these
contributions are quadratically divergent -- independent of the
external momenta). This is the so-called UV/IR mixing problem
\cite{Minwa99}. Similar effects are also known in gauge field models
\cite{Raam01, Haya99, BichlU102}.

A further alternative for describing non-commutative Yang-Mills theory
is given with the Seiberg-Witten map \cite{SeiWi99}. It allows to
connect the non-commutative gauge field with the ordinary gauge field.
Both fields are unified with the ``gauge-equivalent'' gauge
transformation.

In order to carry out perturbative calculations for NCQFT, one can use
a set of modified Feynman rules first formulated in \cite{Filk96} and
elaborated in great detail in \cite{Micu00} for the one-loop and
two-loop approximation of a scalar field model. However, at this
point, one has to stress that the corresponding deformation parameter
$\theta_{\mu\nu}$ characterizing the non-commutativity of space-time
must be restricted to the case $\theta_{0i}=0$. This excludes the
non-locality in the time direction. The only place, where one
encounters time derivatives is the kinetic term of the action and the
usual formalism to get Feynman rules is applicable. The case
$\theta_{0i}\not=0$ is more delicate due to the fact that
non-localities are also present in the time direction. This is really
an obstacle for the naive application of the perturbation theory \`a
la Feynman, where simply the Feynman propagators of the quantum field
models are associated with the Wick contractions. As will be seen
below, this is not the case when time is also involved into
non-locality. Motivated by the works
\cite{Sieb02a,Sieb02b,FischPutz02}, our investigations are devoted to
the discussion of the more general case where non-locality also occurs
in the time direction including NCQFT with a non-vanishing
$\theta_{0i}$. In \cite{FischPutz02}, a compatible time ordering
called \emph{Interaction Point Time Ordered Perturbation Theory
  (IPTOPT)} was proposed in order to evaluate the Gell-Mann-Low
formula in a new way. The corresponding results found in
\cite{Sieb02a,Sieb02b,FischPutz02} are very different from those in
the naive approach treated in the literature. It is also claimed that
with the help of IPTOPT, no unitarity problems arise. As it is shown
in \cite{Bahns02}, the violation of unitarity is due to an improper
definition of QFT on non-commutative space-time. Additionally,
\cite{Rim02} also deals with the unitarity problem at the functional
level.

A further advantage of the proposed IPTOPT is possibly the lack of
the UV/IR mixing problem \cite{FischPutz02,FischPutz03}. In addition,
one also has to mention that non-local interactions become important
in the discussion of ultra-violet finite QFT \cite{Bahns03}.
Therefore, we investigate the perturbation theory of a very general,
non-local interaction of scalar fields allowing the embedding of
NCQFT, UV finite QFT, etc. The main aim of the presented paper is
devoted to find ``simply'' modified Feynman rules for IPTOPT in the
sense of \cite{Filk96, Micu00}, which is also applicable for cases
where non-locality involves time (e.g. $\theta_{0i}=0$). With the help
of these new, modified Feynman rules, it will be demonstrated for some
special cases that the results \cite{Sieb02a, FischPutz02} can be
reproduced very shortly and compactly in a complete new manner.

The paper is organized as follows: In section \ref{sec:nlint}, we
present the main ideas of a non-local, scalar QFT. Section
\ref{sec:Feynmanrules} deals with new Feynman rules for general
non-local interactions. With the help of a simple example, the
``contractor'' will be introduced in analogy to the ordinary covariant
Feynman propagator of local QFT. In section \ref{ssec:cosp}, the
Feynman rules are presented in coordinate space. The explicit
expression for the contractor is elaborated in section
\ref{ssec:contr}. It can be seen immediately that one receives the well
known Feynman propagator in a certain limit. Energy-momentum
conservation at each vertex is discussed in \ref{ssec:pcons}. The
corresponding Feynman-rules in momentum space are given in
\ref{ssec:mosp}. In section \ref{ssec:ncqft}, some specific examples
of NCQFT are studied. The calculation of the tadpole presented in
\ref{sssec:tadp} very powerfully shows the efficiency of the new
Feynman-rules in comparison to the older methods presented in
\cite{FischPutz02}. In section \ref{sssec:t0i0}, the case
$\theta_{0i}=0$ is investigated. Our results confirm those of
\cite{Micu00}. At the end of section \ref{sec:app}, we apply our
approach to the non-local interaction of UV-finite QFT \cite{Bahns03}.
A short discussion is presented in section \ref{sec:disc}. In appendix
\ref{app:NLTO}, one finds the proof of the generalized Wick theorem
for non-locally time-ordered vacuum expectation values.
%
%
%%%%%%%%%%%%%%%%%%%%%%%%%%%%%%%%%%%%%%%%%%%%%%%%%%%
\section{Non-local interactions of general type}
\label{sec:nlint}
%%%%%%%%%%%%%%%%%%%%%%%%%%%%%%%%%%%%%%%%%%%%%%%%%%%

The starting point of NCQFT is based on
\begin{equation}\label{eq:nc0}
        \qcomm{\hat x^\mu}{ \hat x^\nu} = i \theta^{\mu \nu}\qk
        \label{eq:ncx}
\end{equation}
where the hermitian space-time operators are denoted by $\hat x^\mu$.
The antisymmetric deformation parameter $\theta_{\mu\nu}$ is assumed
constant here and has dimension $[{\rm length}]^2$.  In order to deal
with the algebra Eq.~(\ref{eq:nc0}), one introduces the so-called
(Weyl-Moyal) star product defined by \cite{Filk96, Micu00}
\begin{equation}\label{eq:ast}
(f*g)(x) \,\,\equiv\,\,  \left.e^{
        \frac i2 \theta^{\mu\nu}  \partial^\zeta_\mu \partial^\eta_\nu
        }
        f(x+\zeta) g(x+\eta) \right|_{\zeta=\eta=0} \qp 
\end{equation}
In Eq.~(\ref{eq:ast}), $x^\mu$ is now an element of the usual
commutative space-time and $\phi(x)$ is the corresponding scalar
field. The self interaction density of a $\phi^4$-theory would be
modified the following way:
\begin{equation}\label{eq:phi4dens}
  \qVd(z)=\frac \kappa{4!} \phi(z)^4 \,\,\rightarrow\,\,
  \frac \kappa{4!} \phi(z) \ast \phi(z) \ast \phi(z) \ast \phi(z)\qp
\end{equation}
This is related to the interaction $V(z^0)$ in the interaction picture
by $V(z^0)=\int d^3z \qVd(z)$.  For $\theta^{0i} = 0$, it has been
described in \cite{Filk96,Micu00} how to derive Feynman rules and the
outcome showed, that the usual causal Feynman propagator can be used.
However, the interaction vertices are modified by certain phase
factors. The existence of such phase factors leads to planar and
non-planar contributions.  But furthermore, it was pointed out in
\cite{Micu00} that the case $\theta^{0i} \not= 0$ is difficult to
handle due to the fact that the Lagrangian containing star products of
fields consequently also depends on infinitely many time derivatives
acting on fields. Thus, it may be doubted that the Lagrangian
formalism can be applied for $\theta^{0i} \not= 0$ in the usual,
traditional way.

An alternative approach was already followed in
\cite{Sieb02a,FischPutz02}. The calculations carried out there are
based on the Gell-Mann-Low formula
\begin{eqnarray} \label{eq:GellLow}
   \qbra 0 T\{\phi(x_1) \ldots \phi(x_n)\}\qket 0_H
     &=& \nonumber
     \sum_{m=0}^\infty \frac{(-i)^m}{m!}
     \int_{-\infty}^{\infty} dt_{1}
     \int_{-\infty}^{\infty} dt_{2}
     \ldots \int_{-\infty}^{\infty}dt_{m} \times
     \\&&
                \qbra 0 T\{{\phi(x_1) \ldots \phi(x_n) 
                V (t_{1})\ldots V(t_{m})}\}\qket 0_0 
\end{eqnarray}
The subscript $H$ indicates that the operators are given in the
Heisenberg picture and the subscript $0$ refers to the interaction
picture.  Following the derivation of this formula (see \cite{Wein1}
for example), it is clear that time ordering is to be done with
respect to $x_1^0, \ldots,x_n^0$ and $t_1, \ldots ,t_m$, called
\emph{time stamps} from now on.  In order to apply
Eq.~(\ref{eq:GellLow}) for the interaction given by
Eq.~(\ref{eq:phi4dens}), it is helpful to rewrite $\qVd$ in the form
of integral representations \cite{FischPutz02}
\begin{eqnarray}\label{eq:NCVd}
        \qVd(z) &=&\frac\kappa{4!}
        \int\prod_{i=1}^3\left(d^4s_i\frac{d^4l_i}{(2\pi)^4}
        e^{il_i s_i}\right)
        \\
        &&\phi(z-\frac12 \tilde l_1) \phi(z+s_1-\frac12 \tilde l_2)
        \phi(z+s_1+s_2-\frac12 \tilde l_3) \phi(z+s_1+s_2+s_3)
        \qk\nonumber
\end{eqnarray}
where $\tilde a^\nu \equiv a_\mu \theta^{\mu\nu}$.  Time ordering only
involves $z^0$ and not any other time components occurring in the
field operators in Eq.~(\ref{eq:NCVd}).  The advantages of this
representation are that the non-locality can be nicely seen and that
one does not have to care about derivatives, especially time
derivatives, any more as it is the case when explicitly using the star
product in the form of Eq.~(\ref{eq:ast}).

Now, the question is, how to do perturbation theory with interactions
like in Eq.~(\ref{eq:NCVd})? The case of $\theta_{0i}\not=0$ was
already treated in \cite{Sieb02a, FischPutz02} and rules how to do
calculations were also given. But as was also pointed out,
combinatorics was not explicitly treated. Furthermore, it is hard to
see the connection between the rules given in
\cite{Sieb02a,FischPutz02} and ordinary Feynman rules, which should
come out for the case $\theta=0$. Therefore, we have developed
graphical rules in the fashion of Feynman rules for perturbation
theory of non-local interactions of scalar particles of the type
\begin{equation} \label{eq:Vnl}
        V(t) = \int d\qnvec{\lambda}\, v(\qnvec{\lambda},t) 
                \phi(g_1(\qnvec{\lambda},t)) \ldots 
                \phi(g_k(\qnvec{\lambda},t)) 
\end{equation}
in the following section.
$\qnvec{\lambda}=(\lambda_1,\ldots,\lambda_e)$ denotes a set of $e$
real parameters, $v$ is a $C$-function and the $g_i$'s map
$(\qnvec{\lambda},t)$ into a four vector. $d\qnvec\lambda$ simply
abbreviates $d\lambda_1 \ldots d\lambda_e$.

Concerning Eq.~(\ref{eq:NCVd}), this means
\begin{eqnarray*} 
k &=& 4 \qk\\  
\qnvec \lambda &=&(\vec z,l_1,l_2,l_3,s_1,s_2,s_3)\qk\\ 
v(\qnvec{\lambda},t)&=&\kappa /4! \,(2\pi)^{-12}  
\,e^{i (l_1 s_1+l_2 s_2+l_3 s_3)} \qk\\ 
 g_1(\qnvec{\lambda},t)&=&z- \tilde l_1/2 \qk\\ 
g_2(\qnvec{\lambda},t)&=&z+s_1- \tilde l_2/2 \qk\\ 
g_3(\qnvec{\lambda},t)&=&z+s_1+s_2- \tilde l_3/2 \qk\\
g_4(\qnvec{\lambda},t)&=&z+s_1+s_2+s_3\qk
\end{eqnarray*}
with $t=z^0$.  The non-local interaction studied in \cite{Bahns03} was
the main motivation for attacking the problem in such a general way:
\begin{eqnarray}
  V^B_k(z^0) &=& \kappa \,c_k \int d^3z \int d^4a_1\ldots d^4a_k \,\,
  :\phi(z+\zeta a_1)\ldots \phi(z+\zeta a_k):\,\,\times \nonumber\\&&
  \exp{\left\{-\frac12 \sum_{j=1}^k (a_j^\mu)^2\right\}}\,\,
  \delta^4(\frac1k\sum_{j=1}^k a_j)\qk\label{eq:VBahns}
\end{eqnarray}
with $\kappa, c_k, \zeta \in R$. $\kappa$ is the coupling constant and
$c_k$ denotes some normalization constant depending on $k$. $\zeta$
has a physical dimension of a length and should be very small (maybe
in the range of the Planck length). We have explicitly introduced this
parameter, since the limit $\zeta \rightarrow 0$ represents the
corresponding local theory.  We will refer to interactions of this
type as \emph{non-local interactions of Gaussian type} from now on.
Besides the usual normal ordering indicated by the colons, this can be
put into the form of Eq.~(\ref{eq:Vnl}):
\begin{eqnarray*}
 \qnvec \lambda &=&(\vec z,a_1,\ldots a_k)\qk\\
  v(\qnvec{\lambda},t)&=& \kappa \,c_k 
  \exp{\left\{-\frac12 \sum_{j=1}^k (a_j^\mu)^2\right\}}\,\,
  \delta^4(\frac1k\sum_{j=1}^k a_j)\qk\\
  g_j(\qnvec{\lambda},t) &=& z + \zeta a_j\qp
\end{eqnarray*}
with $j=1,\ldots,k$ and $z^0=t$. The effect of normal ordering will be
discussed below.

Before applying the Gell-Mann-Low formula (\ref{eq:GellLow}) to these
interactions, it must be mentioned that $V(t)$ should have a time
dependence according to the interaction picture:
\begin{equation}\label{eq:tinv}
        V(t) = \exp(i H_0 t) V(0) \exp(-i H_0 t) \qp
\end{equation}
Choosing $\phi$ in the interaction picture, this relation can be
easily satisfied by
\begin{eqnarray}
  v(\qnvec{\lambda},t) &=& v(\qnvec{\lambda},0),
  \\
  g_i(\qnvec{\lambda},t)^T
  &=& (g_i(\qnvec{\lambda},0)^1,g_i(\qnvec{\lambda},0)^2, 
     g_i(\qnvec{\lambda},0)^3,g_i(\qnvec{\lambda},0)^0+t).\label{eq:free2}
\end{eqnarray}
The interactions of Eqs.~(\ref{eq:NCVd}) and (\ref{eq:VBahns}) obey
these conditions.  Note that we have adopted the convention
$$
x^\mu = (x,y,z,t) = (x^1,x^2,x^3,x^0)
$$
and we make use of metric defined by
$$
a_\mu = (a_1,a_2,a_3,a_0) \equiv (a^1,a^2,a^3,-a^0)\qp
$$
%
%
%%%%%%%%%%%%%%%%%%%%%%%%%%%%%%%%%%%%%%%%%%%%%%%%%%%
\section{Feynman rules for non-local interactions}
\label{sec:Feynmanrules}
%%%%%%%%%%%%%%%%%%%%%%%%%%%%%%%%%%%%%%%%%%%%%%%%%%%
%
In this section, diagrammatic rules will be given for calculating
\begin{equation}\label{eq:gnm}
        G^n_m(x_1,\ldots,x_n)\equiv 
        \frac{(-i)^m}{m!}\int dt_{n+1}\ldots dt_{N} 
                \qbra 0 {T\{\phi(x_1) \ldots \phi(x_n) 
                V (t_{n+1})\ldots V(t_N)\}}\qket0_0 \qk
\end{equation}
with non-local interactions as given in Eq.~(\ref{eq:Vnl}). $n$ is the
number of external points and $m$ denotes the order of interactions.
$N=n+m$ and the time stamps $t_1,\ldots,t_m$ have been renamed by
$t_{n+1},\ldots,t_{n+m}$.  In order to understand the general case
which will be discussed in appendix \ref{app:NLTO}, we consider in a
first step the simplest non-local interaction of the following form:
$$
V(t) = \int d\qnvec{\lambda}\, v(\qnvec{\lambda},t)
\phi(g_1(\qnvec{\lambda},t)) \phi(g_2(\qnvec{\lambda},t))\qp
$$
Such an interaction could be responsible for
``mass-renormalization'' if one investigates the connected piece of a
``physical'' two-point function
\begin{equation}
  G^2_1(x,y) =\int d\qnvec{\lambda}\, v(\qnvec{\lambda},t) 
  \qbra 0 {T\{
    \phi(x)
    \phi(y)
    \phi(g_1(\qnvec{\lambda},t))
    \phi(g_2(\qnvec{\lambda},t))
  \}}\qket0_0 \qp \label{eq:g21}
\end{equation}
As pointed out in \cite{Sieb02a,Sieb02b,FischPutz02,Rim02}, one
emphasizes here that the time ordering involves only the times $x^0$,
$y^0$ and the time stamp of the interaction part due to the unitarity
problem which is discussed in \cite{Bahns02,Rim02}. Thus, time
ordering is not done with respect to $g_i(\qnvec \lambda, t)^0$, but
simply $t$.

Abbreviating $g_1(\qnvec{\lambda},t)=z+a$ and
$g_2(\qnvec{\lambda},t)=z+b$, with $t$ only occurring in $z^0=t$ ,
Eq.~(\ref{eq:free2}) is automatically satisfied. We are now able to
evaluate Eq.~(\ref{eq:g21}) in terms of free field commutators of
$\phi^+(x)$ and $\phi^-(x)$. In pushing annihilation operators
$\phi^+(x)$ to the right or creation operators $\phi^-(y)$ to the left
in using
\begin{eqnarray}
  \phi(x) &=& \phi^+(x) + \phi^-(x)\qk \\
  \qcomm{\phi^+(x)} {\phi^-(y)} &\equiv& \Delta^+(x,y)\qk
\end{eqnarray}
where we use the following conventions
\begin{eqnarray}
  \phi^+(x) &=& (2 \pi)^{-\frac32} \int d^3p 
  \frac{e^{i\qons px}}{\sqrt{2\qom p}} \,\,a(\vec p)\qk\label{eq:phip}\\
  \phi^-(x) &=& (2 \pi)^{-\frac32} \int d^3p 
  \frac{e^{-i\qons px}}{\sqrt{2\qom p}} \,\,a^\dagger(\vec p)\qk
  \label{eq:phim}\\
  \qcomm{a^\dagger(\vec p)}{a(\vec q)} &=& \delta^3(\vec p -\vec q) 
  \label{eq:acomm}\qk
\end{eqnarray}
where $\qom p \equiv \sqrt{\vec p^2 +m^2}$ and $q^\sigma\equiv(\vec
q,\sigma \qom q)^T$, the connected part of the vacuum expectation
value can now be written as
\begin{eqnarray}
&&  \qbra 0 {T\{
    \phi(x)
    \phi(y)
    \phi(z+a)
    \phi(z+b)
  \}}\qket0_0^{con}
=\nonumber\\
&&\quad \theta(x^0-y^0)\theta(y^0-z^0)
\left[
\Delta^+(x,z+a)\Delta^+(y,z+b)+\Delta^+(x,z+b)\Delta^+(y,z+a)
\right]\nonumber\\
&&+\,\theta(y^0-x^0)\theta(x^0-z^0)
\left[
\Delta^+(x,z+a)\Delta^+(y,z+b)+\Delta^+(x,z+b)\Delta^+(y,z+a)
\right]\nonumber\\
&&+\,\theta(x^0-z^0)\theta(z^0-y^0)
\left[
\Delta^+(x,z+a)\Delta^+(z+b,y)+\Delta^+(x,z+b)\Delta^+(z+a,y)
\right]\nonumber\\
&&+\,\theta(y^0-z^0)\theta(z^0-x^0)
\left[
\Delta^+(z+a,x)\Delta^+(y,z+b)+\Delta^+(z+b,x)\Delta^+(y,z+a)
\right] \label{eq:G21}\\
&&+\,\theta(z^0-x^0)\theta(x^0-y^0)
\left[
\Delta^+(z+a,x)\Delta^+(z+b,y)+\Delta^+(z+b,x)\Delta^+(z+a,y)
\right]\nonumber\\
&&+\,\theta(z^0-y^0)\theta(y^0-x^0)
\left[
\Delta^+(z+a,x)\Delta^+(z+b,y)+\Delta^+(z+b,x)\Delta^+(z+a,y)
\right] \nonumber \qp
\end{eqnarray}
The first (second) summands appearing in the square brackets above
will be referred to as the uncrossed (crossed) terms. The crossed
terms can be obtained from the uncrossed ones by simply exchanging $x
\leftrightarrow y$ or $a \leftrightarrow b$.  Now, let us pick out the
uncrossed terms containing $\Delta^+(y,z+b)$ of Eq.~(\ref{eq:G21}):
\begin{eqnarray*}
&\Delta^+(y,z+b)&
\left[
 \theta(x^0-y^0)\theta(y^0-z^0)\Delta^+(x,z+a)
+\theta(y^0-x^0)\theta(x^0-z^0)\Delta^+(x,z+a) + \right.\\
&&\left.
\theta(y^0-z^0)\theta(z^0-x^0)\Delta^+(z+a,x)
\right]\qp
\end{eqnarray*}
For $(x^0\not=y^0\not=t)$ \footnote{The contributions where some time
  stamps are the same can be neglected.}  , one can now apply a trick:
multiply the first term in square brackets with $\theta(x^0-z^0)$, the
second with $\theta(y^0-z^0)$, the third with $\theta(y^0-x^0)$, and
add within the square brackets
$\theta(y^0-z^0)\theta(z^0-x^0)\theta(x^0-y^0)\Delta^+(z+a,x)$, which
is $0$. Thus, one does not alter the result and gets
$$
\theta(y^0-z^0) \Delta^+(y,z+b) \left\{
  \theta(x^0-z^0)\Delta^+(x,z+a) + \theta(z^0-x^0)\Delta^+(z+a,x)
\right\} \qp
$$
The expression in curly brackets clearly reminds of the usual
Feynman propagator
\begin{equation}
  -i \Delta_F(x,x') = 
  \theta(x^0-{x'}^0) \Delta^+(x,x')
+ \theta({x'}^0-x^0) \Delta^+(x',x) \qp
\end{equation}
For $a=0$, which would be the case for a local field theory, it
reduces to $\Delta_F$. Therefore, we will define
\begin{equation}\label{eq:nlp}
  -i \Delta(x,t;x',t')  \equiv 
  \theta(t-t')\Delta^+(x,x')
+ \theta(t'-t)\Delta^+(x',x)
\end{equation}
and call $\Delta$ the \emph{contractor}. The name is created in
analogy to the usual Wick-contractions of commutative local
perturbation theory in the sense of Feynman.  The semicolon just
visualizes the connection between four vectors $x$ or $x'$ and $t$ and
$t'$, respectively.  Treating the remaining terms of
Eq.~(\ref{eq:G21}) in the same way, $G^2_1$ can now be rewritten as
\begin{eqnarray}
  G^2_1(x,y) = - \int d\qnvec{\lambda}\, v(\qnvec{\lambda},t) 
   &\times&\left[
   \Delta(x,x^0;g_1(\qnvec{\lambda},t),t)
   \Delta(y,y^0;g_2(\qnvec{\lambda},t),t)\right.\\
&&+ \left.\Delta(x,x^0;g_2(\qnvec{\lambda},t),t)
   \Delta(y,y^0;g_1(\qnvec{\lambda},t),t)
\right] \qp\nonumber
\end{eqnarray}
This example will be further discussed in section \ref{sec:app}.
Before, the diagrammatic rules for the general case should be given.
%
%%%%%%%%%%%%%%%%%%%%%%%%%%%%%%%%%%%%%%%%%%%%%%%%%%%
\subsection{Coordinate space rules}
\label{ssec:cosp}
Again, we refer to appendix \ref{app:NLTO}, where the general case
including beside the connected parts also tadpole contributions and
disconnected diagrams has been treated.  In order to calculate
$G^n_m(x_1,\ldots,x_n)$, one has to apply the following diagrammatic
algorithm:

\begin{itemize}
\item Draw $n$ points and label them with the external coordinates
  $x_1,\ldots,x_n$. Their time stamps are $x_1^0,\ldots,x_n^0$,
  respectively.
\item Draw $m$ circles and label them with the parameters $\qnvec
  \lambda_1,\ldots,\qnvec\lambda_m$ and time stamps
  $t_{n+1},\ldots,t_{n+m}$.
\item Draw $k$ points into each circle and label them with
  $g_1,\ldots,g_k$.
\item For each possibility of connecting two points pairwise by a
  line, so that each point is connected to exactly one line, draw a
  diagram with points and circles as given above.
\item For each line connecting two points with coordinates and time
  stamp $x$, $t$ and $x'$, $t'$, respectively, write down a contractor
\begin{equation}
        -i \Delta(x,t;x',t') \nonumber\qk 
\end{equation}
if the points do not belong to the same circle.  If they belong to the
same circle, write down
$$
\textnormal{either}\quad \Delta^+(x,x') \quad \textnormal{or}\quad
\Delta^+(x',x)\qk
$$
depending on whether $\phi(x)$ stands left of $\phi(x')$ within the
interaction $V(t)$ or vice versa.  External points already carry the
coordinates as label and the corresponding time stamp is simply the
$0$th component of that label. For points within circles, the time
stamp $t$ is simply the time stamp of the circle. The coordinate $x$
of such a point is given by the label $\qnvec\lambda$ and $t$ of the
circle and the label $g_j$ of the point as $ x=g_j(\qnvec\lambda,t)\qp
$
\item For each circle labeled with $\qnvec\lambda_i$ and $t_{n+i}$,
  perform an integration according to
  $$
  (-i)\int dt_{n+i}\, d\qnvec\lambda_i v(\qnvec\lambda_i,t)\qp
  $$
\item Sum up the contributions of all diagrams.
\end{itemize}

The rules given above are considerably more complicated than the usual
Feynman rules of the corresponding local field theory. But using this
diagrammatic formalism, Eq.~(\ref{eq:Gmnt}) is much more comfortable
to handle than the usual algorithm of commuting out all creation and
annihilation fields.  It should also be mentioned that for each
diagram with $m$ identical vertices, there are $m!$ diagrams which
only differ by a rearrangement of the vertices.  This fact has already
been implemented in the rules to cancel the factor $1/m!$ of the
Gell-Mann-Low formula (\ref{eq:GellLow}). Consequently, one must
include diagrams differing by a rearrangement of vertices exactly
once.

Further simplifications might be possible when permuting the labels
$g_i$ of a given circle among each other. But this will in general
depend on the interaction itself and will be studied below for a more
special type of interactions. Here it is a good point to comment on
the prescription for calculating tadpole contributions. The rules
given above mean that for a tadpole loop between points with
coordinates $x$ and $x'$, one has to include either $\Delta^+(x,x')$
or $\Delta^+(x',x)$, depending on the definition of the interaction.
If the time stamp of the circle is $t$, the contractor for these
coordinates would give
$$
-i \Delta(x,t;x',t) = \theta(0)\left[ \Delta^+(x,x') +
  \Delta^+(x',x)\right]
$$
which clearly is not what one needs for the general case. But in a
local field theory, $\Delta^+(x,x)$ is to be taken for tadpoles loops.
Thus, by defining $\theta(0)\equiv 1/2$, one does not have to treat
tadpole contributions exceptionally, but can also use the contractor
$-i \Delta(x,t;x,t) = \Delta^+(x,x)$ in local field theories.
%
%%%%%%%%%%%%%%%%%%%%%%%%%%%%%%%%%%%%%%%%%%%%%%%
\subsection{Calculation of the contractor}
\label{ssec:contr}
%%%%%%%%%%%%%%%%%%%%%%%%%%%%%%%%%%%%%%%%%%%%%%%
%
In order to evaluate the contractor, one uses
Eqs.~(\ref{eq:phip}),(\ref{eq:phim}) and (\ref{eq:acomm}) to get
\begin{equation}
        \Delta^+(x,x')=
        \qcomm{\phi^+(x)}{\phi^-(x')} = 
        (2 \pi)^{-3} \int \frac{d^3p}{2\qom p} e^{i\qons p(x-x')}
        \qp
\end{equation}
Expressing the $\theta$-function as
\begin{equation}\label{eq:theta}
\theta(t) = \frac{-1}{2 \pi i} \int_{-\infty}^{\infty} ds
  \frac{e^{-ist}}{s+i\epsilon} \qk
\end{equation}
one obtains
\begin{eqnarray*}
        &&\theta(t-t')\Delta^+(x,x')
        \quad=\\
        &&\frac i{(2\pi)^4} \int \frac{d^3p ds}{2 \qom p}
        e^{i \qons p (x-x')-is(x^0-{x'}^0)}
        \frac{
        e^{is(x^0-t-({x'}^0-t'))}
        }
        {s+i\epsilon}\quad=\\
        &&
        \frac i{(2\pi)^4} \int \frac{d^4p}{2 \qom p}
        e^{i p (x-x')+ip^0(x^0-t-({x'}^0-t'))}
        \frac{e^{-i\qom p(x^0-t-({x'}^0-t'))}}
        {p^0-\qom p + i \epsilon}\qp
\end{eqnarray*}
The last line was obtained by the transformation $s \equiv p^0-\qom
p$.  The other term of Eq.~(\ref{eq:nlp}) can be obtained by simply
exchanging $(x,t)$ and $(x',t')$, and additionally one transforms $p
\rightarrow -p$. The obtained expression only differs in the last
fraction which is
$$
\frac{e^{i\qom p(x^0-t-({x'}^0-t'))}} {-p^0-\qom p + i \epsilon}\qp
$$
Thus, one finally gets
\begin{eqnarray}
        \Delta(x,t;x',t') &=& 
        \frac {1}{(2\pi)^4}\int d^4q
        \frac{e^{iq(x-x')+iq^0(x^0-t-({x'}^0-t'))}}
        {q^2+m^2-i\epsilon}\\
        &&
        \left[\cos(\omega_{\vec q} (x^0-t-({x'}^0-t')))
        -\frac{i q^0}{\omega_{\vec q}} 
        \sin(\omega_{\vec q} (x^0-t-({x'}^0-t')))
        \right] \nonumber
        \qp
\end{eqnarray}
For theories, where the time stamps are always identical with the time
components of field arguments, say $t=x^0$ and $t'={x'}^0$, this
reduces to the usual Feynman propagator
$$
\Delta_F(x,x') = \frac {1}{(2\pi)^4}\int d^4q \frac{e^{iq(x-x')}}
{q^2+m^2-i\epsilon}\qp
$$

%
%
%%%%%%%%%%%%%%%%%%%%%%%%%%%%%%%%%%%%%%%%%%%%%%%%%%%%%%%%%
\subsection{Energy-momentum conservation} 
\label{ssec:pcons}
\begin{figure}
  \centering \epsfig{file=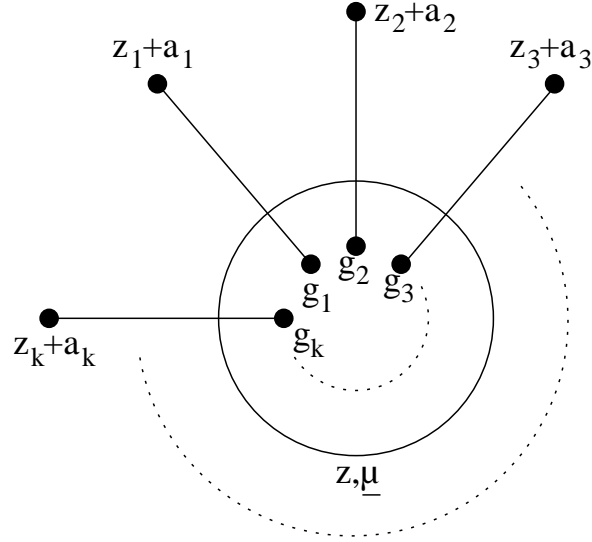,scale=0.8}
  \caption{Vertex of an interaction with $k$ fields. The points out of the circle are considered to be either external or part of other vertices. Their time stamps are  $z_1^0,\ldots,z_k^0\qp$ The time stamp of the circle is $z_0\qp$}\label{fig:vertk}
\end{figure}

%%%%%%%%%%%%%%%%%%%%%%%%%%%%%%%%%%%%%%%%%%%%%%%%%%%%%%%%%
%
Now we present the calculation of diagrams as shown in
Fig.~(\ref{fig:vertk}), for example, using translationally invariant
interactions.  Fig.~(\ref{fig:vertk}) consists of one interaction
vertex containing $k$ points labeled $g_i$ with $i=1,\ldots, k$. These
points all carry the same time stamp. The $k$ ``external'' points are
labeled $z_i+a_i$ all carrying different time stamps $z_i^0$
($i=1,\ldots, k$). Actually, a typical external point would carry time
stamp $z_i^0+a_i^0$. But the intension of this calculation is to work
out Feynman rules in more detail. The ``external'' points are thus
kept more general in order to be also usable as internal points of
larger diagrams containing Fig.~(\ref{fig:vertk}).  The very general
interaction defined by Eq.~(\ref{eq:Vnl}) is now specialized in order
to be translationally invariant:
\begin{eqnarray}
\qnvec \lambda &=& (\vec z,\qnvec \mu) \label{eq:pcons1}\qk\\
v(\qnvec \lambda,t) &=& w(\qnvec \mu)\qk\label{eq:pcons2}\\
g_i(\qnvec \lambda,t) &=& z + h_i(\qnvec \mu) \label{eq:pcons3}\qk
\end{eqnarray}
where bold symbols denote three-vectors and $z^0 \equiv t$.
$\qnvec{\mu}=(\mu_1,\ldots,\mu_{e-3})$ denotes a set of $e-3$ real
parameters, $w$ is a $C$-function and the $h_i$'s map $\qnvec{\mu}$
into a four vector. $d\qnvec\mu$ simply abbreviates $d\mu_1 \ldots
d\mu_{e-3}$.  The very general interaction defined by
Eq.~(\ref{eq:Vnl}) is thus specialized slightly:
\begin{equation}
  V(z^0) =
  \int d^3z \int d\qnvec{\mu}\,\, w(\qnvec{\mu}) \,\,\,\,
                \phi(z+h_1(\qnvec{\mu})) \ldots 
                \phi(z+h_k(\qnvec{\mu}))  \qp
\label{eq:Vnlti}
\end{equation}
This way, translation invariance is manifested:
\begin{equation}
  e^{-i \epsilon  P_{(0)} } \,\,V(z^0) \,\,e^{i \epsilon  P_{(0)} } 
  = V(z^0+\epsilon^0)\qk
\end{equation}
where $P_{(0)}$ denotes the free four-momentum operator generating
translations. $\epsilon$ represents the constant translation in
coordinate space. This way, Eq.~(\ref{eq:tinv}) is automatically
satisfied.

Concerning the non-commutative interaction given by
Eq.~(\ref{eq:NCVd}), the above specialization means
\begin{eqnarray*}
k&=&4\qk \\ 
\qnvec \mu &=& (l_1,l_2,l_3,s_1,s_2,s_3)\qk\\  
w(\qnvec \mu)&=&\kappa /(4! (2\pi)^4) 
\,\exp{(i (l_1 s_1+l_2 s_2+l_3 s_3))}\qk\\ 
h_1(\qnvec{\mu},t)&=&- \tilde l_1 /2\qk\\
h_2(\qnvec{\mu},t)&=&s_1-\tilde l_2 /2\qk\\
h_3(\qnvec{\mu},t)&=&s_1+s_2- \tilde l_3/2\qk\\
h_4(\qnvec{\mu},t)&=&s_1+s_2+s_3\qk\\
\int d\qnvec\mu &=& \int d^3z \prod_{i=1}^3\int ds_i\int dl_i \qp
\end{eqnarray*}
Now we are able to write down the contribution of the diagram shown in
Fig.~(\ref{fig:vertk}):
\begin{eqnarray}
  && G_I(z_1+a_1,z_1^0;\ldots;z_k+a_k,z_k^0) \quad\equiv\quad
  -i \int d\qnvec\mu \,d^4z\,\,
  w(\qnvec \mu) \times\\
  &&(-i\Delta(z_1+a_1,z_1^0;z+h_{1}(\qnvec \mu),z^0)
  \ldots
  (-i\Delta(z_k+a_k,z_k^0;z+h_{k}(\qnvec \mu),z^0)
  \qk\nonumber
\end{eqnarray}
where $I$ stands for the identity permutation of $S^k$, and $S^k$ is
the group of permutations of the integers $\{1,\ldots, k\}$ (see also
\cite{Hamermesh} for more details on permutations). $G_I$ is only one
certain contribution. On the whole, there are $k!$ terms similar to
$G_I$ but only differing in the way the external points are connected
to the internal ones. Denoting an arbitrary permutation by $Q \in
S^k$, one can write these terms as
\begin{eqnarray}
  && G_Q(z_1+a_1,z_1^0;\ldots;z_k+a_k,z_k^0) \quad\equiv\quad
  -i \int d\qnvec\mu \,d^4z\,\,
  w(\qnvec \mu) \times\\
  &&(-i\Delta(z_1+a_1,z_1^0;z+h_{Q_1}(\qnvec \mu),z^0)
  \ldots
  (-i\Delta(z_k+a_k,z_k^0;z+h_{Q_k}(\qnvec \mu),z^0)
  \qp\nonumber
\end{eqnarray}
$G_Q$ gives the vertex where each point $z_i+a_i$ is attached to
$g_{Q_i}(\qnvec \lambda,t)=z+h_{Q_i}(\qnvec \mu)$.  The total
contribution of these terms is given by
\begin{equation}
  G(z_1+a_1,z_1^0;\ldots;z_k+a_k,z_k^0) \quad\equiv\quad
  \sum_{Q \in S^k} G_Q(z_1+a_1,z_1^0;\ldots;z_k+a_k,z_k^0) \qp
\end{equation}
All contractors connecting external and internal points are of the
following form:
\begin{equation}
  \Delta(z+a,z^0; z'+a',{z'}^0)
  =
  \frac1{(2 \pi)^4}\int d^4q
        \frac{e^{iq(z-z')}}
        {q^2+m^2-i\epsilon}
        \sum_{\sigma \in \{1,-1\}}
          \frac{\qom q + \sigma q^0}{2 \qom q} e^{i q^\sigma (a-a')}\qp
\end{equation}
It is remarkable that $z$ and $z'$ only occur in the first
exponential. This expresses translation invariance for time ordered
perturbation theory of non-local interactions also involving time.
Integrating over $z$ immediately yields energy-momentum conservation
and one gets
\begin{eqnarray}
  &&G(z_1+a_1,z_1^0;\ldots;z_k+a_k,z_k^0) \nonumber\\
  &&=  \frac{(-i)^{k+1}}{(2\pi)^{4k}} 
  \sum_{Q \in S^k}\int d\qnvec \mu \,\,w(\qnvec \mu)
\int d^4q_1\ldots d^4q_k \,\,(2\pi)^4 \delta^4(q_1+\ldots+q_k) 
\times \nonumber\\
&&\prod_{i=1}^k \left(\frac{e^{i q_i z_i}}{q_i^2+m^2-i \epsilon}
  \sum_{\sigma_i}
  \frac{\qom {q_i} + \sigma_i q_i^0}{2 \qom {q_i}}
  e^{i q_i^{\sigma_i}(a_i-h_{Q_i})}\right) \nonumber\\
&&=
\frac{(-i)^{k+1}}{(2\pi)^{4k}} \int d^4q_1\ldots d^4q_k 
\,\,(2\pi)^4 \delta^4(q_1+\ldots+q_k) \prod_{i=1}^k 
\left(\frac{e^{i q_i z_i}}{q_i^2+m^2-i \epsilon}\right) \times 
\\
&&
\sum_{\sigma_1,\ldots,\sigma_k} 
\prod_{i=1}^k 
\left(\frac{\qom {q_i} + \sigma_i q_i^0}{2 \qom {q_i}}
e^{i q_i^{\sigma_i}a_i}\right)
\chi(q_1^{\sigma_1},\ldots,q_k^{\sigma_k})
 \qk\nonumber
\end{eqnarray}
with
\begin{equation}\label{eq:phase}
\chi(q_1^{\sigma_1},\ldots,q_k^{\sigma_k}) \equiv 
\int d\qnvec \mu \,\,w(\qnvec \mu)
\sum_{Q \in S^k} \exp{\left(-i\sum_{i=1}^k 
q_{Q_i}^{\sigma_{Q_i}} h_{i}(\qnvec \mu)\right)}\qp
\end{equation}
Of course, the calculated vertex is much more complicated than what
one knows from local field theories. But the positive aspect is given
by the fact that the details of non-locality only occur in the factor
$\chi$. For the case of NCQFT, it contains the phase factors leading
to planar and non-planar contributions. $\chi$ can be said to be the
central quantity of time ordered perturbation theory for non-local
interactions.

Corresponding to this result, it is now straight forward to extract
the momentum space Feynman rules for non-local interactions.
%
%%%%%%%%%%%%%%%%%%%%%%%%%%%%%%%%%%%%%%%%%%%%%%%%%%%%%%%%%
\subsection{Momentum space rules} 
\label{ssec:mosp}
%%%%%%%%%%%%%%%%%%%%%%%%%%%%%%%%%%%%%%%%%%%%%%%%%%%%%%%%%
%
Usually, one is interested in the Fourier transform of
Eq.~(\ref{eq:GellLow}). Correspondingly, one has to evaluate
\footnote{The superscript $con,nt$ indicates that we restrict
  ourselves to connected diagrams without tadpoles. The discussion of
  tadpoles will be treated as an example in section \ref{sec:app}.}
\begin{eqnarray}
G^{con,nt}(p_1,\ldots,p_n) &\equiv&
   \prod_{i=1}^n\left(\int d^4x_i\,e^{i p_i x_i}\right) 
\qbra 0 T\{\phi(x_1) \ldots \phi(x_n)\}\qket 0_H^{con,nt} \nonumber\\
&=& (2\pi)^4 \delta^4(\sum_{i=1}^n p_i)
\prod_{i=1}^{n-1}\left(\int d^4z_i\,e^{i p_i z_i}\right) 
\qbra 0 T\{\phi(z_1) \ldots \phi(z_{n-1})\phi(0)\}\qket 0_H^{con,nt}
\nonumber\\
&\equiv& \label{eq:gtranc}
(2\pi)^4 \delta^4(\sum_{i=1}^n p_i) 
\,\,G^{con,nt}_{tranc}(p_1,\ldots,p_{n-1})\qp
\end{eqnarray}
This task can be simplified using the rules of the last section.  The
$\delta$-function in Eq.~(\ref{eq:gtranc}) represents the total
momentum conservation under the assumption that all momenta are
flowing into the graph.  The main result of the calculation of the
last section is the fact that summing up diagrams, which only differ
by the way how the contractors are attached to points within fixed
circles, can easily be done by including the factor $\chi$ of
Eq.~(\ref{eq:phase}) for each vertex.  But, using $\chi$, one has to
take care of two things:
\begin{itemize}
\item When calculating $\sum_Q G_Q$, it was assumed that all momenta
  are flowing into the vertex. Thus, one has to insert $(-p)^\sigma$
  into $\chi$ if $p$ is flowing out of the vertex.
\item Furthermore, for certain diagrams, one would get double
  counting. Thus one has to include a symmetry factor $1/S$ (see also
  \cite{Sieb02a}). It must be taken the same as for common Feynman
  diagrams.
\end{itemize}
Before presenting the final rules, we will briefly repeat topological
considerations. A general graph is characterized by a certain number
of internal lines $I$ and by the number $V$ of interaction vertices of
a given type. The number of independent loops are denoted by $L$,
which is
\begin{equation}
  L=I-V+1 \qp
\end{equation} 
With the above momentum assignment that all momenta are incoming, we
are now able to state the Feynman rules for the calculation of
expression (\ref{eq:gtranc}):
\begin{enumerate}
\item Draw all possible momentum space Feynman diagrams having $E=n$
  external lines.
\item Carefully label each line with four momentum including its flow
  and make use of the conservation of four momentum at each vertex.
  Due to definition (\ref{eq:gtranc}), the external lines are labeled
  with momenta $p_1,\ldots,p_n$ with the convention that the $p_i$'s
  are incoming. Also assign a variable $\sigma_i$ to each line.
\item For each line, one has to include a factor
  $$
  \frac{-i}{q^2+m^2-i\epsilon}\ \frac{\qom {q} + \sigma q^0}{2 \qom
    {q}}\qk
  $$
  where $q$ and $\sigma$ now represent the labels of the
  corresponding line.
\item For each vertex, write down a factor
  $$-i\,\chi(\ldots)$$
  with the rule to insert $(\pm
  q_i)^{\sigma_i}=(\pm \vec q_i,\pm \sigma_i \qom{q_i})^T$ into $\chi$
  for each line (at the vertex) labeled $q_i$, $\sigma_i$; the ``$+$''
  sign for momenta flowing into the vertex and ``$-$'' otherwise. Due
  to the symmetry of $\chi$ concerning permutations of arguments, the
  order of arguments is not relevant.
\item Include the symmetry factor
  $$1/S\qp$$
\item Assure momentum conservation by a factor
  $$(2 \pi)^4 \delta^4(p_1+\ldots+p_n)\qp$$
\item Integrate over the $L$ independent loop momenta, which are not
  fixed by energy-momentum conservation and multiply by $(2
  \pi)^{-4L}$. Sum over all $\sigma$'s.
\item Sum up all diagrams in the usual sense.
\end{enumerate}
Now one is ready to apply the result of the rather general treatment
of time ordered perturbation theory valid for a large class of
non-local interactions to special cases.
%
%%%%%%%%%%%%%%%%%%%%%%%%%%%%%%%%%%%%%%%%%%%%%%%%%%%
\section{Applications}
\label{sec:app}
%%%%%%%%%%%%%%%%%%%%%%%%%%%%%%%%%%%%%%%%%%%%%%%%%%%
%
%
%%%%%%%%%%%%%%%%%%%%%%%%%%%%%%%%%%%%%%%%%%%%%%%%%%%
\subsection{Non-commutative interactions}
\label{ssec:ncqft}
%%%%%%%%%%%%%%%%%%%%%%%%%%%%%%%%%%%%%%%%%%%%%%%%%%%
%
In this section, we study interactions of the following type:
\begin{equation}
  V_k(z^0) =\frac \kappa{k!} \int d^3z\,\, (\phi(z))^{\ast k} \qk
\end{equation}
where $\ast$ in the power indicates that the star product is to be
used between the $k$ fields.  For $k=4$, this is just the interaction
density given in Eq.~(\ref{eq:phi4dens}) integrated over $d^3z$. In
general, one can write
\begin{equation}
  V_k(z^0) =\frac \kappa{k!}\int d^3z\,\,\prod_{i=1}^{k-1}
  \left( 
    \int\frac{dl_i\,ds_i}{(2\pi)^4} \,e^{i l_i s_i}
    \phi(z-\frac12\tilde l_i+\sum_{j=1}^{i-1} s_j)
  \right)
  \phi(z+\sum_{j=1}^{k-1} s_j) \qp
\end{equation}
These interactions are of the type as specified in
Eqs.~(\ref{eq:pcons1}-\ref{eq:pcons3}) and thus obey momentum
conservation.  The $h_i$'s of Eq.~(\ref{eq:phase}) are then
\begin{eqnarray}
  h_i(\qnvec \mu) &=& -\frac12 \tilde l_i + \sum_{j=1}^{i-1} s_j 
\quad \textnormal{ for } 1\leq i < k\qk\\
  h_k(\qnvec \mu) &=& \sum_{j=1}^{k-1} s_j \qk
\end{eqnarray}
where $\qnvec \mu = (l_1,s_1,\ldots,l_{k-1},s_{k-1})$.  The factor
$\chi$ of Eq.~(\ref{eq:phase}) can then be evaluated for the
interactions $V_k$ (abbreviating $p_i\equiv q_i^{\sigma_i}$):
\begin{eqnarray}
  \chi_k(p_1,\ldots,p_k) &=& \frac \kappa{k!}\sum_{Q \in S^k} 
  \exp{\left(-i \sum_{i<j}p_{Q_i}\wedge p_{Q_j}\right)}
\end{eqnarray}
with
$$
a \wedge b \,\, \equiv \,\, \frac12 \theta_{\mu\nu} \,a^\mu b^\nu
\,\,=\,\,-\frac12 \,a \,\tilde b \qp
$$
$\chi_2$ is simply
\begin{equation}
  1/\kappa\,\,\chi_2(p_1,p_2) =  \cos(p_1 \wedge p_2)
\end{equation} and 
\begin{eqnarray}
  3/\kappa\,\,\chi_3(p_1,p_2,p_3) &=&
  \cos(p_1\wedge p_2 +p_1\wedge p_3 +p_2\wedge p_3)\nonumber\\
&+&\cos(p_1\wedge p_2 +p_1\wedge p_3 -p_2\wedge p_3) \\
&+&\cos(p_1\wedge p_2 -p_1\wedge p_3 -p_2\wedge p_3)\nonumber\qp
\end{eqnarray}
Unfortunately, $\chi_4$ becomes really lengthy:
\begin{eqnarray}
\frac{12}\kappa\chi_4(p_1,p_2,p_3,p_4) &=&
   \cos(p_1\wedge p_2 - p_1\wedge p_3 - p_1\wedge p_4 
- p_2\wedge p_3 - p_2\wedge p_4 - p_3\wedge p_4)\nonumber\\ 
&+&\cos(p_1\wedge p_2 + p_1\wedge p_3 - p_1\wedge p_4 
- p_2\wedge p_3 - p_2\wedge p_4 - p_3\wedge p_4)\nonumber\\ 
&+&\cos(p_1\wedge p_2 + p_1\wedge p_3 + p_1\wedge p_4 
- p_2\wedge p_3 - p_2\wedge p_4 - p_3\wedge p_4)\nonumber\\ 
&+&\cos(p_1\wedge p_2 + p_1\wedge p_3 - p_1\wedge p_4 
+ p_2\wedge p_3 - p_2\wedge p_4 - p_3\wedge p_4)\nonumber\\ 
&+&\cos(p_1\wedge p_2 + p_1\wedge p_3 + p_1\wedge p_4 
+ p_2\wedge p_3 - p_2\wedge p_4 - p_3\wedge p_4)\nonumber\\ 
&+&\cos(p_1\wedge p_2 + p_1\wedge p_3 + p_1\wedge p_4 
+ p_2\wedge p_3 + p_2\wedge p_4 - p_3\wedge p_4)\\ 
&+&\cos(p_1\wedge p_2 - p_1\wedge p_3 - p_1\wedge p_4 
- p_2\wedge p_3 - p_2\wedge p_4 + p_3\wedge p_4)\nonumber\\ 
&+&\cos(p_1\wedge p_2 - p_1\wedge p_3 + p_1\wedge p_4 
- p_2\wedge p_3 - p_2\wedge p_4 + p_3\wedge p_4)\nonumber\\ 
&+&\cos(p_1\wedge p_2 + p_1\wedge p_3 + p_1\wedge p_4 
- p_2\wedge p_3 - p_2\wedge p_4 + p_3\wedge p_4)\nonumber\\
&+&\cos(p_1\wedge p_2 - p_1\wedge p_3 + p_1\wedge p_4 
- p_2\wedge p_3 + p_2\wedge p_4 + p_3\wedge p_4)\nonumber\\ 
&+&\cos(p_1\wedge p_2 + p_1\wedge p_3 + p_1\wedge p_4 
- p_2\wedge p_3 + p_2\wedge p_4 + p_3\wedge p_4)\nonumber\\ 
&+&\cos(p_1\wedge p_2 + p_1\wedge p_3 + p_1\wedge p_4 
+ p_2\wedge p_3 + p_2\wedge p_4 + p_3\wedge p_4)\nonumber\qp
\end{eqnarray}
The term
\begin{equation}
\prod_{j=1}^3\left(\int d^4x_j\,e^{i p_j x_j}\right)
\int dt \qbra 0 T\{\phi(x_1)\ldots\phi(x_k) V_k(t)\}\qket 0_0^{con}
\equiv
G^k_1(p_1,\ldots,p_k)^{con}
\end{equation}
is interesting because it can be compared to the result in
\cite{Sieb02a} for $k=3$.  Using the momentum space rules one
immediately gets
\begin{eqnarray}\label{eq:phi3vert}
  G^k_1(p_1,\ldots,p_k)^{con}&=&
   (2 \pi)^{4}  \delta^4(p_1+\ldots+p_k) \times
\\&&
\sum_{\sigma_1,\ldots,\sigma_k}
  \prod_{j=1}^k\frac{-i }{2 \qom{p_j} 
    (\qom{p_j}-\sigma_j p_j^0-i\epsilon)} 
  \,\chi(p_1^{\sigma_1},\ldots,p_k^{\sigma_k})\nonumber\qp
\end{eqnarray}
We also made use of the identity
$$
\frac 1{q^2+m^2-i \epsilon} = \frac1{2 \qom q} \left( \frac1{\qom q
    -\sigma q^0-i \epsilon} + \frac1{\qom q +\sigma q^0-i \epsilon}
\right)
$$
which holds for infinitesimal $\epsilon$ and $\sigma=\pm 1$.
Eq.~(\ref{eq:phi3vert}) with $k=3$ is identical to the corresponding
expression given in \cite{Sieb02a}.

The result for $k=2$ is remarkable. It can be written as
\begin{equation}
G^2_1(p,q)^{con}
 \,\,=\,\,  
 {\kappa}  \,\,\delta(p+q)
\,\,\frac{(-i)^3 {(2 \pi)^4}}{(p^2+m^2-i \epsilon)^2} \,\, 
\label{eq:vert2}
\frac{
\qom p^2+{p^0}^2+(\qom p^2-{p^0}^2)\cos(p^+\wedge p^-)}
{2\qom {p}^2}
 \qp
\end{equation}
It is surprising that this term contains a phase factor which only
vanishes for $\theta_{0i}=0$. On-shell, it does not have any unusual
effect since $\qom p^2=({p^0})^2$. But off-shell it might be used as
an extra counterterm within loops or other internal lines. If one
starts with the naive Lagrangian approach, quadratic terms do not
contain phases since
$$
\int d^4x\,\, (f\ast g)(x) = \int d^4x\,\,f(x)\,g(x)\qp
$$
This relation was taken from \cite{Micu00}, where it was further
argued that this is the reason why the free theory of NCQFT is the
same as in the commutative case and only the interaction is modified.
Apparently, this might seem problematic for our work since we used
$H_0$ of the commutative case as free Hamiltonian. But this problem
can be resolved the following way: Suppose, there was some
non-commutative full Hamiltonian $H_\ast$. Then one can simply extract
the interaction $V$ by definition:
$$
V \equiv H_\ast - H_0 \qk
$$
which means that perturbation theory is done with respect to $H_0$
and not any other part of $H_\ast$ which reminds us of $H_0$ but
containing some $\ast$-products. Of course, we cannot say that this
way of doing perturbation theory will be the best one but at least, it
is possible to do some perturbative calculations within this
framework. The result presented in Eq.~(\ref{eq:vert2}) shows that the
naive Lagrangian approach and time ordered perturbation theory for
non-local interactions worked out here are not equivalent when
non-locality also involves time ($\theta_{0i}\not=0$).
%
%%%%%%%%%%%%%%%%%%%%%%%%%%%%%%%%%%%%%%%%%%%%%%%%
\subsubsection{Tadpole}
\label{sssec:tadp}
%%%%%%%%%%%%%%%%%%%%%%%%%%%%%%%%%%%%%%%%%%%%%%%%
Before dealing with the special case $\theta_{0i}=0$, the calculation
of a tadpole according to the coordinate space rules should be given,
namely
$$
G_{tp}(x,y) \equiv -i\int dz^0\,\,\qbra 0 T\{\phi(x) \phi(y)
V_4(z^0)\}\qket 0_H^{con} \qp
$$
In this case, all connected contributions contain exactly one line
starting and ending in the circle representing the interaction, which
means that one only has diagrams containing tadpoles. Thus we cannot
use the momentum space rules which have not been worked out for
tadpole contributions, but we have to take the somewhat lengthier
coordinate space rules.  We simply consider this as an exercise and
check of our approach: using normal ordered interactions, there would
not be tadpole contributions. Therefore, we do not see any directly
physical relevance of such contributions.
\begin{figure}
  \centering \epsfig{file=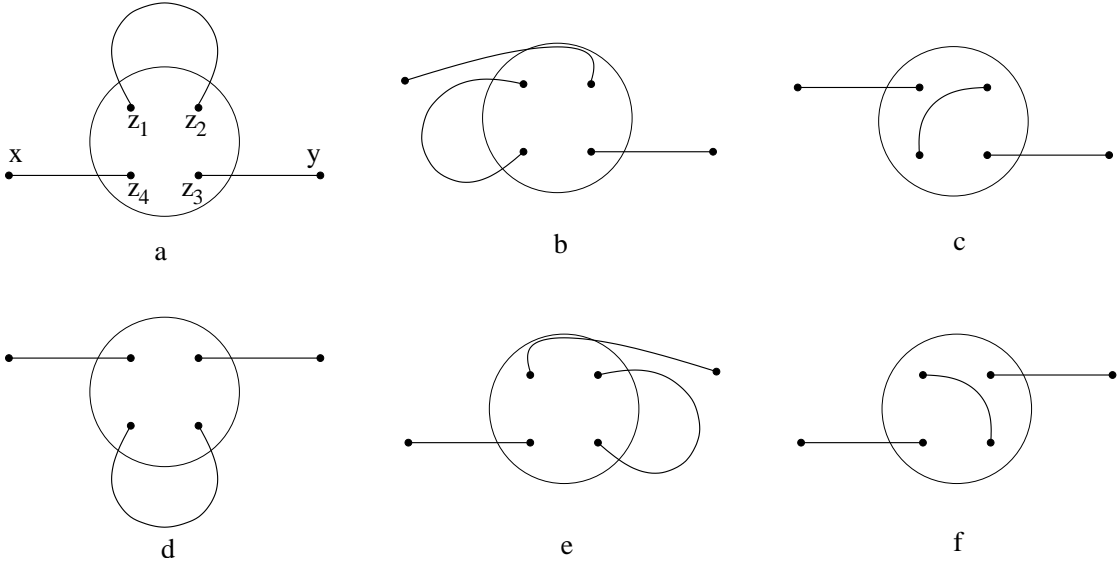,scale=0.9}
  \caption{This figure shows half of the diagrams contributing to 
    $G_{tp}(x,y)$. The crossed diagrams which are obtained by 
    exchanging $x$ and $y$ have been omitted. Diagrams b - f are 
    considered to be labeled like a. We have used the abbreviations 
    $z_1=z-\frac12 \tilde l_1$, $z_2=z-\frac12 \tilde l_2+s_1$,
    $z_3=z-\frac12 \tilde l_3+s_1+s_2$, $z_4=z + s_1+s_2+s_3\qp$ }
  \label{fig:tadp}
\end{figure}
The result is obtained by calculating the diagrams shown in
Fig.~(\ref{fig:tadp}) and the corresponding crossed diagrams which are
obtained by simply exchanging $x$ and $y$. Diagram c and its crossed
one give
$$
-i \int dz \prod_{i=1}^3 \left(\frac {d^4l_i d^4s_i}{(2\pi)^4} e^{i
    l_i s_i}\right) \left[ \Delta^+(z_2,z_4) (-i
  \Delta(x,x^0;z_1,z^0)) (-i \Delta(y,y^0;z_3,z^0)) + x
  \leftrightarrow y \right]\qk
$$
with $z_1,\ldots,z_4$ as described in Fig.~(\ref{fig:tadp}).
Summing up all 12 diagrams and carrying out the trivial integrations,
one arrives at
\begin{eqnarray}
        G_{tp}(x,y) &=&
        \frac{i \kappa}{4!}\int \frac{d^4q}{(2\pi)^4 \qom{q}^2}
        \frac{d^3k}{(2\pi)^3 2 \qom{k}}\,\,
        e^{iq(x-y)}
        \nonumber\\
        &&
        \left[
        \frac{
                2+\cos(q^+ \tilde{k}^+)
        }
        {(q^0-\qom{q} + i \epsilon)^2}
        +
        \frac{
                2+\cos(q^- \tilde{k}^+)
        }
        {(q^0+\qom{q} - i \epsilon)^2}
        \right.
        \label{eq:tadp}\\&&
        \left.
        -
        \cos(\frac12 q^- \tilde{q^+}) \frac{
                3
                +\cos(q^+ \tilde{k}^+)
                +\cos((q^+ - q^-)\tilde{k}^+)
                +\cos(q^- \tilde{k}^+)
        }
        {({q^0}^2-\qom{q}^2 + i \epsilon)}
        \right]\nonumber \qp
\end{eqnarray}
This is consistent with the result of \cite{FischPutz02}. Diagrams c
and f (and their crossed counterparts) shown in Fig.~(\ref{fig:tadp})
are responsible for the contributions containing phase factors at the
pole of order two at $q^0 \rightarrow \qom q$ of the integrand of
Eq.~(\ref{eq:tadp}). Furthermore, it is interesting that the momentum
$q$ of external lines always occurs as $q^\pm$ in the
$\cos$-functions. For $\theta_{0i}=0$, it would not make any
difference whether we write $q^\pm \tilde {k^\pm}$ or $q \tilde {k}$,
but for $\theta_{0i}\not=0$, it definitely does. This new aspect also
effects the UV/IR-mixing problem discussed in \cite{FischPutz03}.

%
%%%%%%%%%%%%%%%%%%%%%%%%%%%%%%%%%%%%%%%%%%%%%%%%
\subsubsection{$\theta_{0i}=0$}
\label{sssec:t0i0}
%%%%%%%%%%%%%%%%%%%%%%%%%%%%%%%%%%%%%%%%%%%%%%%%
Let us now turn to the special case of $\theta_{0i}=0$. Rewriting
$$
a \wedge b = \frac12 a_i b_j \theta^{ij} \qk
$$
one sees that $\chi_k$ does not depend on any time component of its
arguments, and one can for example replace the arguments $(\pm
q)^\sigma$ by the off-shell momentum $\pm q$. Thus it is independent
of the $\sigma$'s originating from the momentum space rules, and the
sums over $\sigma$'s can be evaluated independently:
$$
\sum_{\sigma \in \{1,-1\}} \frac{\qom {q} + \sigma q^0}{2 \qom {q}}
= 1\qp
$$
The only thing differing from common field theory are the factors
$\chi_k$ associated with the vertices. The conservation of four
momentum leads to the following simplifications for $\theta_{0i}=0$:
\begin{eqnarray}
  \chi_2(p_1,p_2) &=& \kappa \qk\\
  \chi_3(p_1,p_2,p_3) &=& \kappa \,\,\cos(p_1\wedge p_2) \qk\\
  \chi_4(p_1,p_2,p_3,p_4) &=& 
  \kappa/3
  \left[\cos(p_1 \wedge p_2)\cos(p_3 \wedge p_4)\right.\\
  &&\left.\quad+
    \cos(p_1 \wedge p_3)\cos(p_2 \wedge p_4)+
    \cos(p_1 \wedge p_4)\cos(p_2 \wedge p_3) 
  \right]\nonumber\qp
\end{eqnarray}
The resulting $\theta_{0i}=0$-rules for $k=2,4$ are identical with the
ones given in \cite{Micu00}.
%
%
%%%%%%%%%%%%%%%%%%%%%%%%%%%%%%%%%%%%%%%%%%%%%%%%%%%
\subsection{Non-local interactions of Gaussian type}
\label{ssec:fredl}
%%%%%%%%%%%%%%%%%%%%%%%%%%%%%%%%%%%%%%%%%%%%%%%%%%%
%
It can easily be seen that the interaction of Eq.~(\ref{eq:VBahns})
proposed in \cite{Bahns03} is translationally invariant in the sense
of Eqs.~(\ref{eq:pcons1}-\ref{eq:pcons3}). So the momentum space rules
are applicable. In order to specify perturbation theory further, one
simply has to calculate $\chi$ of Eq.~(\ref{eq:phase}):
\begin{eqnarray}
\chi(p_1,\ldots,p_k) &=& 
\kappa c_k \int d^4a_1 \ldots d^4a_k\,\,\times\\&&\nonumber
\exp{\left\{-\frac12 \sum_{j=1}^k (a_j^\mu)^2\right\}}\,\,
  \delta^4(\frac1k\sum_{j=1}^k a_j)\,\,
  \sum_{Q \in S^k} 
  \exp{\left(-i\sum_{j=1}^k p_{Q_j} \,\zeta a_{j}\right)}\qk
\end{eqnarray}
where we have abbreviated $p_i \equiv q_{i}^{\sigma_{i}}$. A lengthy
but standard calculation involving a multiple Gaussian integral then
yields
\begin{equation}
  \chi(p_1,\ldots,p_k) =
  \kappa \,\,(2 \pi)^{2k-2}(k-1)!\, c_k 
  \,\,e^{-k \,\zeta^2\left(\qmean{(p^\mu)^2}
      -\left(\qmean{p^\mu}\right)^2\right)}\qk \label{eq:chiBahns}
\end{equation}
with $\qmean A$ representing the mean value
$$
\qmean A \equiv \frac1k\sum_{j=1}^k A_j \qp
$$
We have written $\chi$ in such a statistical manner in order to see
that the exponent in Eq.~(\ref{eq:chiBahns}) is always negative. Thus,
it seems plausible that the exponential damping caused by $\chi$ makes
the contribution of all diagrams finite. Indeed, $\int dx \,p_n(x)
\,\exp(-x^2)$ is always finite for $p_n$ a polynomial of order $n$.
But one has to keep in mind that $\chi$ never involves time components
$q_i^0$ of the four momenta $q_i$, but $q_i^{\sigma_i}$ instead. So it
is clear that one separately has to check whether the integrations
over time components are finite or not. Concerning the integrations
over the 3-momenta, it seems plausible that they are finite due to the
exponential damping. But investigations concerning these aspects are
still to be carried out within the presented formalism, especially
also for theories involving other types of particles.
%%%%%%%%%%%%%%%%%%%%%%%%%%%%%%%%%%%%%%%%%%%%%%%%%%%%%%%%%
\section{Conclusion} 
\label{sec:disc}
%%%%%%%%%%%%%%%%%%%%%%%%%%%%%%%%%%%%%%%%%%%%%%%%%%%%%%%%%
%
The special case of NCQFT for $\theta_{0i}=0$ can be generalized for
other non-local interactions obeying translation invariance.
Inspecting Eq.~(\ref{eq:phase}), $\chi$ is independent of the time
components of its arguments, if $h_i(\qnvec \mu)^0=0$ for $1\leq i
\leq k$.

Or in other words: \emph{If time is not involved into non-locality,
  non-local interactions with translation invariance can be described
  by the usual Feynman rules with three-momentum-dependent factors to
  be included for each vertex.  }

If time is involved into non-locality of a translationally invariant
interaction, one has to use the momentum space rules given in
\ref{ssec:mosp}. Such rules have in principal already been stated in
\cite{Sieb02a} or \cite{FischPutz02} or . The main difference between
these rules and the ones presented here lies in the fact that we
rearranged the $\theta$-functions arising for time ordering $T$ in a
tricky way before blowing them up using Eq.~(\ref{eq:theta}). Thus, we
do not have to draw separate Feynman diagrams for each possible time
ordering. This makes our rules much simpler to use. The
$\phi^3$-vertex agrees with the result of \cite{Sieb02a} who have
calculated it without making use of the non-local representation of
the interaction. Furthermore we arrive at the same diagrammatic rules
as \cite{Micu00} for $\theta_{0i}=0$. Our results have been derived
not only for NCQFT but for general non-local interactions of the type
$\phi^k$. The interaction $V(t)$ simply has to be invariant under time
translations in the sense of Eq.~(\ref{eq:tinv}). If translation
invariance is not satisfied for the spatial components, one is not
lost because we can refer to the coordinate space rules presented in
\ref{ssec:cosp}. An example for an interaction without translation
invariance concerning spatial components would be the interaction
$$
V(z^0) = \int d^3z \,U(\vec z)\,\, \phi(z)^2\qp
$$
Here, $U$ is an unquantized, external potential constant in time
scattering scalar particles.  Except this simple example, one can
assume that in general, these cases will be more complicated to
handle.

Finally, it must be said that the naive Lagrangian approach cannot be
used when non-locality involves time. This can be seen in the case of
NCQFT: If $\theta_{0i}\not=0$, the Lagrangian contains infinitely many
time derivatives. How can canonical quantization work for these cases?
The failure of the naive Lagrangian approach is not a catastrophe
since an alternative approach has already been followed in
\cite{Sieb02a,FischPutz02} and also in our work. But the problem now
is that this alternative approach only offers the possibility of
studying symmetries perturbatively. When time is not involved into
non-locality, the Lagrangian formalism works in the case of NCQFT and
maybe this is also true for other non-local interactions. 
\\[3mm]

\appendix
%%%%%%%%%%%%%%%%%%%%%%%%%%%%%%%%%%%%%%%%%%%%%%%%%%%
\section{Generalized Wick-theorem for non-locally time ordered 
vacuum expectation values}
\label{app:NLTO}
%%%%%%%%%%%%%%%%%%%%%%%%%%%%%%%%%%%%%%%%%%%%%%%%%%%
%
In this section, the time ordered product of Eq.~(\ref{eq:gnm})
\begin{equation}\label{eq:top1}
  \qbra 0 {T\{\phi(x_1) \ldots \phi(x_n) 
                V (t_{n+1})\ldots V(t_N)\}}\qket0_0
\end{equation}
will be further processed for non-local interactions as given in
Eq.~(\ref{eq:Vnl}).  To deal with the expression given above, it is
quite useful to define the times $t_1,\ldots,t_n$ as
$x_1^0,\ldots,x_n^0$ and
\begin{equation}
        \theta_P(\qnvec t) \equiv 
        \theta(t_{P_1}-t_{P_2})\theta(t_{P_2}-t_{P_3})
        \ldots \theta(t_{P_{N-1}}-t_{P_N})
\end{equation}
where $\qnvec t \equiv (t_1,\ldots,t_N)$ and $P \in S^N$, the group of
permutations of $(1,\ldots, N)$. $P_i$ denotes the integer $P(i)$, the
integer $i$ is mapped to by the permutation $P$ (see also
\cite{Hamermesh} for more details on permutations).  Inserting
Eq.~(\ref{eq:Vnl}), expression~(\ref{eq:top1}) can then be rewritten
as
\begin{eqnarray}
        \int d\qnvec \lambda_1 \ldots d\qnvec \lambda_m 
        v(\qnvec \lambda_1) \ldots
        \sum_{P \in S^N}\theta_P(\qnvec t)
        \qexv{\phi(z_{{\xi^P}_1}) \ldots \phi(z_{{\xi^P}_M}}_0    \qk    
\end{eqnarray}
where $M$ is the number of fields occurring in the vacuum expectation
value and $z_1,\ldots, z_M$ are defined as the arguments put into the
fields from left to right for the time ordering $\theta_P$
corresponding to the identity permutation $P=I$ (which corresponds to
the time ordering $t_1>t_2>\ldots>t_N$). $z_{{\xi^P}_i}$ denotes the
four vector which is put into the $i$th field according to the time
ordering $\theta_P$. Thus, $\xi^P$ is implicitly defined as a
permutation $\in S^M$ for each $P \in S^N$, so that $z_{{\xi^P}_i}$ is
the argument put into the $i$th field (counting from left to right).
%
%%%%%%%%%%%%%%%%%%%%%%%%%%%%%%%%%%%%%%%%%%%%%%%%%%%%%%%%%
%%%%%%%%%%%%%%%%%%%%%%%%%%%%%%%%%%%%%%%%%%%%%%%%%%%%%%%%%
%

The vacuum expectation value in the above Eq.
\begin{equation}\label{eq:vacexp}
        \qexv{\phi(z_{{\xi^P}_1}) \ldots \phi(z_{{\xi^P}_M}}_0
%       \qexv{\phi(z^P_1) \ldots \phi(z^P_M)}_0 
%       \equiv
%       \qexv{\phi(y_1) \ldots \phi(y_M)}_0     
\end{equation}
can be rewritten in terms of commutators by applying the following
algorithm:
\begin{itemize}
\item Substitute every field operator $\phi(y)$ by the sum of its
  annihilation and creation part $\phi^+(y)+\phi^-(y)$.
\item Expand the product into sums of products of $\phi^+$ and
  $\phi^-$.
\item Expand each product by replacing adjacent pairs of operators
  $\phi^+(z_{{\xi^P}_i})\phi^-(z_{{\xi^P}_j})$ by
  $\phi^-(z_{{\xi^P}_j})\phi^+(z_{{\xi^P}_i})
  +D_P({\xi^P}_i,{\xi^P}_j)$, where\footnote{The omission of
    $\theta(j-i)$ would not make any difference at this point since
    for all adjacent operators occurring during application of the
    given algorithm, $i<j$. Below, the purpose of this definition will
    become clear.}
        \begin{equation} \label{eq:commphi}
        D_P({\xi^P}_i,{\xi^P}_j) \equiv 
        \theta(j-i)
        \qcomm{\phi^+(z_{{\xi^P}_i})}{\phi^-(z_{{\xi^P}_j})}\qp
        \end{equation}
        This step has to be repeated till no $\phi^-$ can be found
        right of any $\phi^+$.
      \item Note that $\ldots \phi^+(x)\qket 0 = \qbra 0 \phi^-(x)
        \ldots = 0$.
\end{itemize}
Now, the question is, which kind of terms remains after this
procedure. First of all, it is clear that for $M$ odd, the vacuum
expectation value vanishes since for each summand only a single field
operator would remain. So $M$ even is the more interesting case. Some
interesting properties for this case are:
\begin{itemize}
\item Each summand is a product of $M/2$ functions
  $D_P({\xi^P}_i,{\xi^P}_j)$ with $i<j$ which each integer $\in
  \{1,\ldots,M\}$ occurs exactly once in.
\item Further on, all summands are different from each other, when
  considering $D_P(i,j)$ and $D_P(k,l)$ only identical for $i=k$ and
  $j=l$.
\item Each product of $M/2$ functions $D_P({\xi^P}_i,{\xi^P}_j)$,
  which each integer $\in \{1,\ldots,M\}$ occurs exactly once in, also
  represents a summand. The property $i<j$ of the first item would
  also imply a restriction on the integers $i,j$ occurring in the
  arguments of $D_P$. But due to the apparently useless
  $\theta$-function in Eq.~(\ref{eq:commphi}), terms not obeying this
  restriction are $0$ which makes the restriction obsolete.
\end{itemize}
On the one hand, the first property says that one can write each
summand $S_i$ as
\begin{equation}\label{eq:permsumm}
        S_i = 
        D_P({\xi^P}_{Q_1},{\xi^P}_{Q_2})
        D_P({\xi^P}_{Q_3},{\xi^P}_{Q_4})
        \ldots D_P({\xi^P}_{Q_{M-1}},{\xi^P}_{Q_M})
        %D_P(Q_1,Q_2)D_P(Q_3,Q_4)\ldots D_P(Q_{M-1},Q_M)
        \qk
\end{equation}
where $Q \in S^M$. But since ${\xi^P}_{Q_i}=\xi^P(Q(i))$ and $\xi^P
\in S^M$, one can simply write $Q'_i=Q'(i)=\xi^P(Q(i))$ instead of
${\xi^P}_{Q_i}$.  On the other hand, according to the third property,
it can be said that each term $D_P(Q_1,Q_2)\ldots D_P(Q_{M-1},Q_M)$
with arbitrary $Q$ occurs in the sum.  The vacuum expectation value of
Eq.~(\ref{eq:vacexp}) can be written as
\begin{equation}\label{eq:vacexpand}
        \qexv{\phi(z_{{\xi^P}_1}) \ldots \phi(z_{{\xi^P}_M}}_0
%       \qexv{\phi(z^P_1) \ldots \phi(z^P_M)}_0 
        =
        \frac1{(M/2)!}\sum_{Q\in S^M} 
        D_P(Q_1,Q_2) \ldots D_P(Q_{M-1},Q_M)\qp
\end{equation}
One has to divide by $(M/2)!$ since each product of $M/2$ functions
with 2 arguments is generated by $(M/2)!$ permutations.  Note that
definition~(\ref{eq:commphi}) is equivalent to
\begin{equation}
        D_P(i,j) = \theta({{(\xi^P)}^{-1}}_j-{{(\xi^P)}^{-1}}_i)
        \qcomm{\phi^+(z_i)}{\phi^-(z_j)} \qp
\end{equation}
We have verified Eq.~(\ref{eq:vacexpand}) by carrying out a proof by
induction. But due to its length and the fact that one can understand
the result as described above, we want to skip it.  Combining
Eq.~(\ref{eq:vacexpand}) with the sum over permutations $P$ due to
time ordering, one gets
\begin{eqnarray}
        &&%\frac1 {(M/2)!}
        \sum_{P \in S^N}\theta_P(\qnvec t)
        \sum_{Q\in S^M} 
        D_P(Q_1,Q_2) \ldots D_P(Q_{M-1},Q_M) =\nonumber\\
&&\sum \limits_{Q\in S^M} 
        C(Q_1,Q_2) \ldots C(Q_{M-1},Q_M) \,\,\times
        \label{ex:themonster}\\
        \nonumber\\
        &&\sum \limits_{P \in S^N}\theta_P(\qnvec t) \,\,
        \theta\left({{(\xi^P)}^{-1}}_{Q_2}
          -{{(\xi^P)}^{-1}}_{Q_1}\right)
        \ldots
        \theta\left({{(\xi^P)}^{-1}}_{Q_M}
          -{{(\xi^P)}^{-1}}_{Q_{M-1}}\right)
\nonumber\qk
\end{eqnarray}
where
\begin{equation}
        C(i,j) \equiv \Delta^+(z_i,z_j) \qp
\end{equation}
The (only) nice thing about expression~(\ref{ex:themonster}) is that
the sum over $P$ only depends on $\theta$-functions. For given times
$t_1,\ldots,t_N$, $\theta_P$ now is only survived by one permutation .
The other $\theta$-functions oppose $M/2$ conditions
$$
{{(\xi^P)}^{-1}}_{Q_{2i-1}} < {{(\xi^P)}^{-1}}_{Q_{2i}} \qk
$$
or in other words: In expression~(\ref{eq:vacexp}), the field
evaluated at $z_{Q_{2i-1}}$ had to stand left of the field evaluated
at $z_{Q_{2i}}$.
This condition can easily be rewritten in terms of the time stamps
associated with the corresponding field arguments, provided the two
stamps are not the same. Thus, it is comfortable to define $\tau_i$ as
the mapping
$$
\tau: \{1,\ldots, M\} \,\, \rightarrow \,\, \{t_1,\ldots, t_N\}\qk
$$
where $\tau_i$ is defined to be the time stamp associated with the
field argument $z_i$.  For $\tau_{Q_{2i-1}} \not= \tau_{Q_{2i}}$
\footnote{This is meant to hold only for $\forall i,j \in
  \{1,\ldots,N\}: t_i\not= t_j$. The other cases should be irrelevant:
  Those, which involve time stamps associated with interactions have
  $0$ Lebesgue measure, and those, where the time stamps of two
  external lines are the same, can be omitted anyway.}
$$
\theta\left({{(\xi^P)}^{-1}}_{Q_{2i}}-{{(\xi^P)}^{-1}}_{Q_{2i-1}}\right)
= \theta(\tau_{Q_{2i-1}} - \tau_{Q_{2i}})\qp
$$
$\tau_{Q_{2i-1}} = \tau_{Q_{2i}}$ means that the two time stamps
are the same and come from one interaction, say $D_P(Q_{2i-1},Q_{2i})$
represents a tadpole loop. In this case, the $\theta$-function simply
assures that $\Delta^+(z_{Q_{2i-1}},z_{Q_{2i}})$ is only taken into
account when $\phi(z_{Q_{2i-1}})$ really stands left of
$\phi(z_{Q_{2i}})$ in the interaction.  It is useful to distinguish
between sets of summands $S^M_{nt}$ (no tadpole) where the above
simplification is applicable and where it is not, $S^M_{t}$ (tadpole):
\begin{equation}
\begin{array}{rcl}
        S^M_{nt} &\equiv& \left\{Q \in S^M \right.\left|
        \forall i \in \{1,\ldots,M/2\}:\,\,\tau_{Q_{2i-1}} 
        \not= \tau_{Q_{2i}}\right\}\qk\\\\
        S^M_{t} &\equiv& S^M \setminus S^M_{nt} \qp
\end{array}
\end{equation}
Summarizing the book keeping carried out above, one gets \footnote{In
  the first sum, the factor $1/(2^{M/2})$ arises from using the
  symmetric $\Delta$ which corresponds to blowing up the sum over $Q$
  by replacing $D_P(i,j)$ with $D_P(i,j)+D_P(j,i)$.}
\begin{eqnarray}
\label{eq:Gmnt} 
\label{eq:fineman}
  &&
  \qbra 0 {T\{\phi(x_1) \ldots \phi(x_n) \nonumber 
                V (t_{n+1})\ldots V(t_N)\}}\qket0_0
        \quad=\quad 
        \int d\qnvec \lambda_1 \ldots d\qnvec \lambda_m 
        v(\qnvec \lambda_1) \ldots\\
        &&\left[\frac{1}{(M/2)!2^{M/2}}\sum_{Q\in S^M_{nt}} 
        (-i\Delta(Q_1,Q_2)) \ldots (-i\Delta(Q_{M-1},Q_M))\right.\\
        &&\left.+
        \frac1 {(M/2)!}
        \sum_{P \in S^N}\theta_P(\qnvec t)
        \sum_{Q\in S^M_t} 
        D_P(Q_1,Q_2) \ldots D_P(Q_{M-1},Q_M)\right]\nonumber 
\end{eqnarray}
where $\Delta(i,j)$ is the book keeping version of the contractor for
integer arguments
$$
-i \Delta(i,j) \,\,\equiv \,\, D_P(i,j) + D_P(j,i) \,\,=\,\, -i
\Delta(z_i,\tau_i;z_j,\tau_j)\qp
$$
The second sum still looks complicated, but one can use the
contractor $\Delta$ for all lines but the tadpole lines. For the
latter one, one simply has to insert $\Delta^+$ with arguments
corresponding to the order in the interaction.  Eq.~(\ref{eq:fineman})
now represents the generalized Wick theorem. It is based on
Eq.~(\ref{eq:vacexpand}), which is in principle applicable for an
arbitrary ordering of fields specified by the left hand side.
Combining this relation with the time ordering supposed by the
Gell-Mann-Low formula (\ref{eq:GellLow}), one arrives at
Eq.~(\ref{eq:fineman}), which we refer to the \emph{generalized Wick
  theorem}. It is valid for non-local interactions, in particular,
non-locality may also involve time. This means that $\phi(z_i)$ is not
necessarily ordered with respect to $z_i^0$ but $\tau_i$.  The
ordinary Wick theorem is represented by the special case $\tau_i =
z_i^0$. The contractor then becomes the usual propagator. For this
case, our result agrees with the explicit Wick theorem (see
\cite{Zuber}).  As in this special case, the evaluation of
Eq.~(\ref{eq:fineman}) for the general case can be simplified a lot by
introducing diagrammatic rules. But in principle, it should be clear
how this works and we confine ourselves to presenting the rules in
section \ref{sec:Feynmanrules}.
%
%%%%%%%%%%%%%%%%%%%%%%%%%%%%%%%%%%%%%%%%%
% Bibliagraphy stuff
%%%%%%%%%%%%%%%%%%%%%%%%%%%%%%%%%%%%%%%%%
%
\bibliographystyle{JHEP-2} 
\bibliography{qft}

\end{document}